\UseRawInputEncoding
\RequirePackage{fix-cm}

\documentclass[smallextended]{svjour3}       % onecolumn (second format)
\smartqed  % flush right qed marks, e.g. at end of proof
\usepackage{graphicx}
\usepackage{xcolor}
\usepackage{wrapfig}
\usepackage{subfig}
\usepackage{rotating}
\usepackage{url}
\usepackage{makeidx}
\usepackage{multirow}
\usepackage{stmaryrd} 
\usepackage{setspace}
\usepackage{algpseudocode}
\usepackage{algorithmicx}
\usepackage{amssymb}
\usepackage[justification=centering]{caption}
\usepackage{multirow}
\usepackage[flushleft]{threeparttable}
\usepackage{scrextend}
\usepackage[english]{babel}
\usepackage{blindtext}
\usepackage{adjustbox}
\usepackage{amsmath}
\usepackage{hyphenat} 
\usepackage{float}
\usepackage{braket}
\usepackage{tabularx}
\begin{document}

\title{Threshold Quantum Secret Sharing%\thanks{Grants or other notes
}


\author{Kartick Sutradhar      
}


\institute{Kartick Sutradhar \at
              Indian Institute of Information Technology Sri City \\
              \email{kartick.sutradhar@gmail.com}           %  \\
}

\date{Received: date / Accepted: date}

\maketitle

\begin{abstract}
One crucial and basic method for disclosing a secret to every participant in quantum cryptography is quantum secret sharing. Numerous intricate protocols, including secure multiparty summation, multiplication, sorting, voting, and more, can be designed with it. A quantum secret sharing protocol with a $(t,n)$ threshold approach and modulo $d$—where $t$ and $n$ represent the threshold number of participants and the total number of participants, respectively—was recently discussed by Song {\em et al.}. Kao {\em et al.} notes that without the information of other participants, the secret in Song {\em et al.'s}protocol cannot be reconstructed. We address a protocol that solves this issue in this paper. 

\keywords{Secure Computation \and Quantum Cryptography \and Information Security \and Quantum Secret Sharing}
\end{abstract}
\section{Introduction}
The quantum secret sharing includes a dealer and a group of $n$ participants \cite{sutradhar2021enhanced}. The dealer distributes the shares of a secret among $n$ participants. When the dealer requires to retrieve the original secret, the $t$ (threshold) number of participants will work together to retrieve it \cite{song2017t,sutradhar2020efficient,sutradhar2020generalized}. The quantum  secret sharing can be used in various applications \cite{mashhadi2012analysis,mashhadi2012novel,sun2020toward}, namely, secure multiparty summation \cite{shi2017quantum,sutradhar2021efficient}, multiplication \cite{sutradhar2020hybrid}, comparison, sorting,  voting, etc., as it preserves the secret from getting lost, damaged, or changed \cite{shi2016quantum,shi2016comment,shi2018efficient}. There have been discussed  numerous protocols for sharing a secret in literature \cite{Qin2018Multidimensional,bao2009threshold,sutradhar2021cost,yang2013secret,mashhadi2017new,hillery1999quantum}. There are two approaches followed in quantum secret sharing protocols, namely, $(t,n)$ and $(n,n)$ threshold approaches. The first $(n,n)$ threshold based quantum  secret sharing protocol \cite{hillery1999quantum} was discussed by Hillery {\em et al.} 
\par
An $(t, n)$ threshold quantum based secret sharing protocol with level-$d$ was discussed by Song {\em et al.} in 2017 that used the $CNOT$ operation, $QFT$, generalized Pauli operator, and inverse quantum Fourier transform ($IQFT$) \cite{song2017t}. This protocol includes a dealer and a group of participants. The dealer chooses one participant as a trusted reconstructor and $SHA-1$\cite{eastlake2001us} as the hash algorithm to evaluate the secret hash value. The dealer sends the secret's hash value to a trusted reconstructor, who can recover the secret using a collision attack. Further, the trusted reconstructor cannot reconstruct the original secret from the $IQFT$ operation \cite{CommentKao2018}. The $IQFT$ operation cannot sum up all the states. To recover the original secret, the trusted reconstructor needs other participants' secret information. In 2020, Mashhadi improved the Song {\em et al.'s} protocol \cite{mashhadi2020improvement} by using the $d$-level SUM operation, $QFT$, and $IQFT$. This protocol is efficient but it has high computation and communication costs due to the transmission of $(t-1)$ decoy particle, more  number of $IQFT$ operation, and SUM operation. Moreover, if the reconstructor is corrupted or dishonest, then the threshold number of participants cannot recover the secret in both the Mashhadi’s and Song et al.’s protocols. Hence, in these protocols, the reconstructor must be honest. In addition, similar to the Song {\em et al.'s} protocol, the trusted reconstructor may also recover the secret by performing the collision attack because the dealer sends the secret's hash value to the trusted reconstructor directly. In this paper, we propose a new $d$-level quantum based secret sharing protocol using the $(t,n)$ threshold approach that overcomes the above mentioned problems. We may summarize our contributions as follows.
\begin{itemize}
\item The reconstructor $Bob_1$ can reconstruct the original secret efficiently.
 \item The reconstructor $Bob_1$ cannot reveal the secret by performing the collision attack.
 \item The proposed protocol can also resist the coherent and collective attacks.
 \item The proposed protocol can also detect the eavesdropping by comparing the hash values of the secret even if the reconstructor transmits a fake secret to other participants after recovering the original secret.
\end{itemize}

\section{Preliminaries}
Here, we introduce the Shamir's secret sharing, $QFT$, and $IQFT$, which will be used in our proposed protocol.
\subsection{Shamir's Secret Sharing \cite{shamir1979share}}
This protocol has two phases as discussed below.
\subsubsection{Sharing of Secret}
The dealer creates $n$ shares of the secret using a polynomial $f(x)$ of degree ($t-1$)  and distributes $n$ shares among $n$ participants.
\subsubsection{Reconstruction of Secret}
The threshold number of participants reconstructs the secret as follows.
\begin{equation}\label{equ1}
f(x) = \sum_{v=1}^{t} f(x_v) \prod_{1 \le j \le t, j \neq v} \frac{x_j}{x_j - x_v}
\end{equation}
\subsection{Quantum Fourier Transform ($QFT$) \cite{song2017t,sutradhar2023quantum,sutradhar2021secret,sutradhar20213efficient,sutradhar20244quantum,sutradhar2024smart,challagundla2024privacy,sutradhar2024review,challagundla2024secure,nayaka2024survey}}
The quantum Fourier transform (QFT) is defined as $$QFT: \ket{\alpha}  \rightarrow \frac{1} {\sqrt{d}} \sum_{\beta=0}^{d-1} e^{2\pi i\frac{\alpha}{d}\beta} \ket{\beta}.$$
\subsection*{Inverse Quantum Fourier Transform ($IQFT$) \cite{song2017t,sutradhar2024survey,sutradhar2023secure,sutradhar2024svqcp,sutradhar2022privacy,sutradhar20211efficient,sutradhar2024privacy,sutradhar20244privacy,sutradhar2023quantum}}
The inverse quantum Fourier transform ($IQFT$) is defined as $$IQFT: \ket{\beta}  \rightarrow \frac{1} {\sqrt{d}} \sum_{\alpha=0}^{d-1} e^{-2\pi i\frac{\beta}{d}\alpha} \ket{\alpha}.$$
\section{Review of Song {\em et al.'s} protocol}
Here, we review the Song {\em et al.}'s protocol. In this protocol, the dealer shares a secret $S$ among $n$ participants $\mathcal{B} = \{Bob_1, Bob_2, \dots, Bob_n \}$. From $n$ participants, any one is selected by the dealer as a trusted reconstructor. We may consider here $Bob_1$ as a trusted reconstructor.
\subsection{Distribution of Shares}
The dealer selects an arbitrary polynomial $p(x)$ of degree ($t-1$) such that $p(x) \in \mathbb{Z}_d$, where $\mathbb{Z}_d$ is a finite field. The ($t-1$)-degree polynomial may be defined as $$p(x) = S + a_1x + \dots +a_{t-1}x^{t-1}.$$ A non-zero value $x_i \in \mathbb{Z}_d$ is also selected by the dealer to compute $n$ shares $p(x_i)$. The dealer encodes $p(x_i)'s$ using BB84 and sends the qubit string of $p(x_i)$ through a secure quantum channel to every participant $Bob_i, i=1,2,...,n$. The dealer chooses a hash algorithm $H()$ to determine the hash value $H(S)$ of the secret $S$ and sends this hash value $H(S)$ to the participant $Bob_1$.   
\subsection{Reconstruction of Secret}
The secret is reconstructed by a certain number of participants using the following steps.\\
\\
\textbf{Step 1:} Participant $Bob_1$ (reconstructor) prepares a $t$-qudit particle $\ket{l}_1,\ket{l}_2, \dots, \ket{l}_t$, which contains $m$ qubits, where $m = \lceil \log_{2}^{d} \rceil$. The participant $Bob_1$ applies the $QFT$ on the particle $\ket{l}_1$ that results in the output state $\ket{\varphi_1}$, as follows.
\begin{equation}\label{equ8}
\begin{split}
\ket{\varphi_1} & = \big(QFT\ket{l}_1 \big) \ket{l}_2, \ket{l}_3, \dots, \ket{l}_t\\
& = \Big(\frac{1} {\sqrt{d}} \sum_{u=0}^{d-1} \omega^{0.u} \ket{u}_1 \Big) \ket{l}_2, \ket{l}_3, \dots, \ket{l}_t\\
& = \Big( \frac {1} {\sqrt{d}} \sum_{u=0}^{d-1} \ket{u}_1 \Big) \ket{l}_2, \ket{l}_3, \dots, \ket{l}_t\\
 \end{split}
\end{equation}
\\
\textbf{Step 2:} Participant $Bob_1$ again prepares a $v$-qudit particle $\ket{l}_v$, where $v=2, 3, \dots, t$, which contains $m$ qubits, where $m = \lceil \log_{2}^{d} \rceil$. The participant $Bob_1$ applies the $d$-level $CNOT$ gate \cite{shi2016secure} on the particle $\ket{l}_v$, where $v= 2, 3, \dots, t$. After performing $(t-1)$ number of $CNOT$ gates, the state $\ket{\varphi_1}$ becomes an entangled state $\ket{\varphi_2}$ \cite{zhang2021experimental,hu2019experimental} as follows.
\begin{equation}\label{equ9}
\begin{split}
\ket{\varphi_2} & = (CNOT((QFT\ket{l}_1),\ket{l}_2)) \otimes, \dots, \otimes(CNOT((QFT\ket{l}_1),\ket{l}_t))\\
& = \frac {1} {\sqrt{d}} \sum_{u=0}^{d-1} \ket{u}_1 \ket{u}_2 \ket{u}_3, \dots, \ket{u}_t
 \end{split}
\end{equation}
\\
\textbf{Step 3:} Participant $Bob_1$ sends the particle $\ket{u}_v$ through a secure quantum channel to respective participant $Bob_v$, $v= 2, 3, \dots, t$.\\
\\
\textbf{Step 4:} Each participant $Bob_v$ evaluates the share's shadow $(s_v)$,  $v= 1, 2, \dots, t$, as follows.
\begin{equation}\label{equ10}
s_v  = f(x_v) \prod_{1\leq j\leq t, j\neq v} \frac {x_j} {x_j - x_v} \mod d
\end{equation}
\\
\textbf{Step 5:} The Pauli operator $(U_{0,s_v})$ is applied by each participant $Bob_v$ on their respective private particles $\ket{u}_v$,  $v= 1, 2, \dots, t$, as follows.
\begin{equation}\label{equ11}
 U_{0,s_v} = \sum_{u=0}^{d-1} \omega^{s_v.u} \ket{u}_{v~~v}\bra{u}
\end{equation}
After performing the Pauli operator on each participant particle, the state $\ket{\varphi_2}$ extends as follows:
\begin{equation}\label{equ12}
\begin{split}
\ket{\varphi_3} & = \frac {1} {\sqrt{d}} \sum_{u=0}^{d-1} \omega^{s_1.u} \ket{u}_1 \omega^{s_2.u} \ket{u}_2 \omega^{s_3.u} \ket{u}_3, \dots, \omega^{s_t.u} \ket{u}_t \\
& = \frac {1} {\sqrt{d}} \sum_{u=0}^{d-1} \omega^{(\sum_{v=1}^{t} s_v).u} \ket{u}_1 \ket{u}_2 \ket{u}_3, \dots, \ket{u}_t
 \end{split}
\end{equation}
\\
\textbf{Step 6:} Finally, the participant $Bob_1$ applies the $IQFT$ on his private particle $\ket{u}_1$ and, based on computational basis, measures it to acquire the secret $p(0)'= \sum_{v=1}^{t} s_v \mod d$.
\section{Comments on Song {\em et al.'s} protocol}
Here, we show the incorrectness of the reconstruction phase of the Song {\em et al.'s} protocol. Kao {\em et al.} point out that, without other participants' information, $Bob_1$ can never retrieve the secret. Song {\em et al.} mention that $QFT(\sum_{v=1}^{t} s_v)$ is the qubit of $Bob_1$ in $\ket{\varphi_1}$. The participant $Bob_1$ evaluates $IQFT$ over its particle $QFT(\sum_{v=1}^{t} s_v)$ and measures it on a computational base, where $Bob_1$ retrieves the secret $S' = \sum_{v=1}^{t} s_v$. We have the following observation.
 \begin{equation}
 \begin{split}
 \ket{\phi_1}=&\frac{1}{\sqrt{d}} \sum_{u=0}^{d-1}\omega^{(\sum_{v=1}^{t}s_v).u}\ket{u}_1 \ket{u}_2 \dots \ket{u}_t\\
 & \neq \frac{1}{\sqrt{d}}( \sum_{u=0}^{d-1}\omega^{(\sum_{v=1}^{t}s_v).u}\ket{u}_1) \ket{u}_2 \dots \ket{u}_t\\
 & = QFT(\sum_{v=1}^{t}s_v) \ket{u}_2 \dots \ket{u}_t.
 \end{split}
 \end{equation}    
The secret $S' = \sum_{v=1}^{t} s_v$ cannot be retrieved even when $IQFT$ is performed over the particle $\ket{l}_1$ and measured computationally by $Bob_1$.
\begin{equation}
 \begin{split}
 \ket{\phi_2}=& QFT \otimes I \otimes \dots \otimes I(\frac{1}{\sqrt{d}} \sum_{u=0}^{d-1}\omega^{(\sum_{v=1}^{t}s_v).u} \ket{u}_1 \ket{u}_2 \dots \ket{u}_t)\\
 & = \frac{1}{\sqrt{d}}(\sum_{u=0}^{d-1} IQFT(\omega^{(\sum_{v=1}^{t}s_v).u}\ket{u}_1) \ket{u}_2 \dots \ket{u}_t\\
 & \neq \frac{1}{\sqrt{d}} IQFT(\sum_{u=0}^{d-1}\omega^{(\sum_{v=1}^{t}s_v).u} \ket{u}_1) \ket{u}_2 \dots \ket{u}_t)\\
 & = |\sum_{v=1}^{t}s_v\rangle_1 \ket{u}_2 \dots \ket{u}_t.
 \end{split}
 \end{equation} 
For better understanding of the problem, consider an example, where $d = 3, t=2, n = 4$ and $S =2$. From step $5$ of the reconstruction phase of the Song {\em et al.'s} protocol, we have
   \begin{equation}
 \begin{split}
     \ket{\phi_3} &=\frac{1}{\sqrt{3}}\sum_{u=0}^{2}\omega^{2.u}\ket{uu} \\
     & \neq \frac{1}{\sqrt{3}}( \sum_{u=0}^{2}\omega^{2.u} \ket{u})\ket{u} \\
     &= QFT \ket{2} \ket{u}
   \end{split}
 \end{equation} 
On applying the inverse quantum Fourier transform $IQFT$ over the particle $\ket{u}$, we get
   \begin{equation}
 \begin{split}
 \ket{\phi_4} &= QFT \otimes I \frac{1}{\sqrt{3}}(\ket{00} + \omega^2\ket{11} + \omega \ket{22})\\
 &=\frac{1}{\sqrt{3}}(QFT \ket{0}\ket{0} + \omega ^2 IQFT \ket{1} \ket{1} + \omega IQFT \ket{2} \ket{2})\\
& = \frac{1}{3}((\ket{0} + \ket{1} + \ket{2})\ket{0} + \omega^2(\ket{0} + \omega^{-1}\ket{1} +\omega^{-2}\ket{2})\ket{1} + \omega(\ket{0} + \omega^{-2}\ket{1}+ \omega^{-1}\ket{2}) \ket{2})\\
 & = \frac{1}{3}(\ket{0}(\ket{0} +\omega^{2}\ket{1} + \omega \ket{2}) + \ket{1}(\ket{0} + \omega \ket{1} +\omega^{2}\ket{2})  + \ket{2}(\ket{0} + \ket{1} + \ket{2})).
 \end{split}
 \end{equation} 
 The result to the equation comes out as $\ket{0}, \ket{1}$ or $\ket{2}$, not accurately $\ket{2}$.
 \subsection*{Attack on Song {\em et al.'s} Protocol}
The dealer chooses $Bob_1$ as a trusted reconstructor in the Song {\em et al.} protocol, and the hash algorithm $SHA-1$ to evaluate the secret's hash value. After computing the hash value, the dealer transfers this hash value through a secure quantum channel to $Bob_1$. From this hash value, $Bob_1$ can easily reveal the secret by performing the collision attack.

\section{Proposed quantum secret sharing protocol}
Here, we propose a new quantum  secret sharing protocol that has $(t, n)$ threshold and $d$-level. The distribution of the shares and the reconstruction of secret are its two main phases, as discussed below.
\subsection{Distribution of share}
The dealer selects an arbitrary $(t-1)$-degree polynomial  $p(x) \in \mathbb{Z}_d$, $\mathbb{Z}_d$ is a finite field, as follows: $$p(x) = S + a_1x + \dots +a_{t-1}x^{t-1}.$$ The dealer selects a non-zero value $x_i \in \mathbb{Z}_d$ to compute $n$ shares $p(x_i)$, encodes $p(x_i)s$ using BB84 and sends the qubit string of $p(x_i)$ via a secure quantum channel to every participant $Bob_i, i=1,2,..,n$. Then, the dealer chooses a hash algorithm to determine the secret hash value $\mathcal{H}(S)$. After computing $\mathcal{H}(S)$, the dealer shares it using a polynomial $h(x)=\mathcal{H}(S) + \gamma_1x + \gamma_2x^2 + \dots + \gamma_{t-1}x^{t-1}$ among $n$ participants. Participant $Bob_i$ only learns the share $h(x_i)$, $i= 1, 2, \dots, n$.
\subsection{Reconstruction of the secret}
Let $\mathcal{B}=\{Bob_1, Bob_2, \dots, Bob_t\}$ be a  qualified subset of $t$ participants. The dealer chooses a reconstructor participant from the qualified subset. In this phase, the dealer chooses $Bob_1$ as a reconstructor participant that recovers the secret and the secret hash value using the following steps:\\
\\
\textbf{Step 1:} Reconstructor $Bob_1$ prepares $t$ qudit particle $\ket{l}_1,\ket{l}_2, \dots, \ket{l}_t$, which contains $m$ qubits, $m = \lceil \log_{2}^{d} \rceil$. The participant $Bob_1$ applies the $QFT$ \cite{shi2016secure} on the particle $\ket{l}_1$. The output state $\ket{\varphi_1}$ is computed as follows.
\begin{equation}\label{equ8}
\begin{split}
\ket{\varphi_1} & = \big(QFT\ket{l}_1 \big) \ket{l}_2, \ket{l}_3, \dots, \ket{l}_t\\
& = \Big(\frac{1} {\sqrt{d}} \sum_{u=0}^{d-1} \omega^{0.u} \ket{u}_1 \Big) \ket{l}_2, \ket{l}_3, \dots, \ket{l}_t\\
& = \Big( \frac {1} {\sqrt{d}} \sum_{u=0}^{d-1} \ket{u}_1 \Big) \ket{l}_2, \ket{l}_3, \dots, \ket{l}_t\\
 \end{split}
\end{equation}
\\
\textbf{Step 2:} The participant $Bob_1$ prepares $v$ qudit particle $\ket{l}_v$,  $v=2, 3, \dots, t$ and this particle contains $m$ qubits, $m = \lceil \log_{2}^{d} \rceil$. $Bob_1$ performs $d$-level $CNOT$ gate on the particle $\ket{l}_v$, where $v= 2, 3, \dots, t$. After performing $(t-1)$ $CNOT$ gates, the state $\ket{\varphi_1}$ becomes an entangled state $\ket{\varphi_2}$ \cite{zhang2021experimental,hu2019experimental}  as follows.
\begin{equation}\label{equ9}
\begin{split}
\ket{\varphi_2} & = (CNOT((QFT\ket{l}_1),\ket{l}_2)) \otimes, \dots, \otimes(CNOT((QFT\ket{l}_1),\ket{l}_t))\\
& = \frac {1} {\sqrt{d}} \sum_{u=0}^{d-1} \ket{u}_1 \ket{u}_2 \ket{u}_3, \dots, \ket{u}_t
 \end{split}
\end{equation}
\\
\textbf{Step 3:} $Bob_1$ sends the particle $\ket{u}_v$, $v= 2, 3, \dots, t$, to respective $Bob_v$ participants through a secure quantum channel.\\
\\
\textbf{Step 4:} Each participant $Bob_v$ evaluates the share's shadow $(s_v)$, $v= 1, 2, \dots, t$.
\begin{equation}\label{equ10}
s_v  = f(x_v) \prod_{1\leq j\leq t, j\neq v} \frac {x_j} {x_j - x_v} \mod d
\end{equation}
\\
\textbf{Step 5:} The Pauli operator $(U_{0,s_v})$ applied by each participant $Bob_v$ on his private particle $\ket{u}_v$,  $v= 1, 2, \dots, t$.
\begin{equation}\label{equ11}
 U_{0,s_v} = \sum_{u=0}^{d-1} \omega^{s_v.u} \ket{u}_{v~~v}\bra{u}
\end{equation}
After performing the Pauli operator on each participant particle, the state $\ket{\varphi_2}$ extends as follows:
\begin{equation}\label{equ12}
\begin{split}
\ket{\varphi_3} & = \frac {1} {\sqrt{d}} \sum_{u=0}^{d-1} \omega^{s_1.u} \ket{u}_1 \omega^{s_2.u} \ket{u}_2 \omega^{s_3.u} \ket{u}_3, \dots, \omega^{s_t.u} \ket{u}_t \\
& = \frac {1} {\sqrt{d}} \sum_{u=0}^{d-1} \omega^{(\sum_{v=1}^{t} s_v).u} \ket{u}_1 \ket{u}_2 \ket{u}_3, \dots, \ket{u}_t
 \end{split}
\end{equation}
\\
\textbf{Step 6:} Each participant $Bob_v$ applies the $IQFT$ on his private particle $\ket{u}_v$ and measures the result of $IQFT$. After measuring, each participant $Bob_v$ broadcasts the result of measurement.\\ 
\\
\textbf{Step 7:} Each participant $Bob_v$ computes the secret $p(0)'= \sum_{v=1}^{t} s_v \mod d$ by adding the measurement results.\\
\\
\textbf{Step 8:} Finally, all seven steps discussed above are again performed by the threshold number of participants $t$ to reconstruct the secret hash value. The secret hash value $h(0)'=\sum_{r=1}^{t}~g_r~mod~d$ is reconstructed by the participant $Bob_1$, where $g_r$ represents the hash value share's shadow. The participant $Bob_1$ uses the hash algorithm $SHA-1$ to determine the hash value $\mathcal{H}(p(0)')$ and matches it with the secret' hash value $h(0)'$. If $(\mathcal{H}(p(0)')=h(0)')$, then the participant $Bob_1$ perceives that the threshold number of participants have executed the protocol honestly; otherwise, $Bob_1$ believes that the one or more corrupt participants have executed the protocol.
\section{Security Analysis}
In this section, we discuss the collision, coherent, and collective attacks, which can be resisted by the proposed protocol. 
\subsection{Collision attack} 
An attacker uses the hash algorithm attack to generate the same secret hash value for two inputs in this attack. In the Song {\em et al.'s} \cite{song2017t} and Mashhadi's \cite{mashhadi2020improvement} protocols, the $Bob_1$ can execute the collision attack to get the secret because the dealer sends the secret's hash value to $Bob_1$ and hence it is not secure against the collision attack. Our protocol is secure against the collision attack because the dealer determines the secret hash value and shares this value among $n$ participants. So, the reconstructor participant $Bob_1$ has no knowledge about the hash value and hence he is unable to execute the collision attack.
\subsection{Coherent attack}
In this attack, an attacker creates an independent ancillary particle $\ket{w}$ and intercepts every participant's particle $\ket{l}_v$ by jointly interacting with every qudit of participant $Bob_v, v=1,2,\dots, t$. On every participant's particle $\ket{l}_v$, the attacker conducts the measurement process in computational basis. The attacker just gets $l$ with $\frac{1}{d}$ probability from this calculation of particle $\ket{l}_v$. However, $l$ does not hold any valuable data about the share's shadow. Only the interacting particle $\ket{l}_v$ is known to the attacker in this case. As a result, the attacker cannot get the share's shadow from the coherent attack.
\subsection{Collective attack}
In a collective attack, an attacker communicates with each qudit by creating an individual ancillary particle and performing a measure all of the ancillary qudits at the same time to obtain the share's shadow. Every qudit of participant $Bob_v, v=1,2,\dots, t$ is interacted with by an individual ancillary particle $\ket{w}$ created by the attacker. After communicating, the attacker obtains the particle $\ket{l}_v$ and conducts a joint calculation procedure in the computational basis to reveal the share's shadow. Since the particle $\ket{l}_v$ does not hold any valuable data about the share's shadow, the attacker cannot obtain any information about it from this joint calculation.
\section{Performance Analysis}
Here, we analyze  the performance of the proposed protocol and compare with that of the Song {\em et al.'s} \cite{song2017t}, and Mashhadi's \cite{mashhadi2020improvement} protocols in terms of the security and cost. The  Song em et al.'s protocol \cite{song2017t} requires one $QFT$ operation, $t$ unitary operations, two hash operations, one $IQFT$ operation, one measure operations, and transmit $(t-1)$ message particles. This protocol is not efficient because the $IQFT$ cannot recover the original secret. The Mashhadi's protocol \cite{mashhadi2020improvement} needs one $QFT$ operation, $t$ unitary operations, two hash operations, $t$ number of $IQFT$ operations, $(t-1)$ SUM operations, $t$ measure operations, and transmit $(t-1)$ message particles with $(t-1)$ decoy particles. However, our protocol requires one $QFT$ operation, $t$ unitary operations, two hash operations, $(t-1)$ $IQFT$ operation, $(t-1)$ measure operations, and transmit $(t-1)$ number of message particles. Moreover, the Mashhadi's protocol uses the SUM operation, more number of $IQFT$ operation, and transmission of $(t-1)$ decoy particles; whereas, our protocol uses $CNOT$ gate, less number of $IQFT$ operation, and no transmission of the decoy particles. Hence, it has high cost as compared to our protocol. In addition, the proposed protocol is more cost effective, efficient, and secure as compared to the Song {\em et al.'s} \cite{song2017t}, and Mashhadi's \cite{mashhadi2020improvement} protocols. Table \ref{Table1} shows the comparison of these protocols.
\begin{table}[ht]
\centering
\caption{Comparison of security and cost}
\begin{tabular}{|l|c|c|c|}
\hline
\multicolumn{1}{|c|}{\textbf{Performance Parameter}} & \textbf{Song {\em et al.}} \cite{song2017t} & \textbf{Mashhadi \cite{mashhadi2020improvement}} & \textbf{Proposed} \\ \hline
SUM operation                                & -          & $(t-1)$  & -          \\ \hline
Measure operation                            & 1          & $t$      & $(t-1)$    \\ \hline
Unitary operation                            & $t$        & $t$      & $t$        \\ \hline
Decoy particle                               & -          & $(t-1)$  & -          \\ \hline
Message particle                             & $(t-1)$    & $(t-1)$  & $(t-1)$    \\ \hline
QFT operation                                & 1          & 1        & 1          \\ \hline
IQFT operation                               & 1          & $t$      & $(t-1)$    \\ \hline
Prevention of collision attack                             & No         & No       & Yes        \\ \hline
Prevention of coherent attack                              & No         & No       & Yes        \\ \hline
Prevention of collective attack                            & No         & No       & Yes        \\ \hline
\end{tabular}
\label{Table1}
\end{table}
\section{Conclusion}
In this paper, we have discussed a new $(t,n)$ threshold protocol for quantum secret sharing in which the reconstructor can reconstruct the original secret efficiently. This protocol can execute the threshold number of participants without any trusted reconstructor participant. Further, the secret hash value and the secret are unknown to the reconstructor participant and he cannot execute the collision attack, but can correctly execute the proposed protocol. The proposed protocol can also resist the coherent and collective attacks.
%\bibliography{biblio.bib}

\begin{thebibliography}{10}

\bibitem{sutradhar2021enhanced}
Kartick Sutradhar and Hari Om.
\newblock Enhanced (t, n) threshold d-level quantum secret sharing.
\newblock {\em Scientific Reports}, 11(1):17083, 2021.

\bibitem{song2017t}
Xiu-Li Song, Yan-Bing Liu, Hong-Yao Deng, and Yong-Gang Xiao.
\newblock (t, n) threshold d-level quantum secret sharing.
\newblock {\em Scientific reports}, 7(1):6366, 2017.

\bibitem{sutradhar2020efficient}
Kartick Sutradhar and Hari Om.
\newblock Efficient quantum secret sharing without a trusted player.
\newblock {\em Quantum Information Processing}, 19(2):1--15, 2020.

\bibitem{sutradhar2020generalized}
Kartick Sutradhar and Hari Om.
\newblock A generalized quantum protocol for secure multiparty summation.
\newblock {\em IEEE Transactions on Circuits and Systems II: Express Briefs}, 67(12):2978--2982, 2020.

\bibitem{mashhadi2012analysis}
Samaneh Mashhadi.
\newblock Analysis of frame attack on hsu et al.’s non-repudiable threshold multi-proxy multi-signature scheme with shared verification.
\newblock {\em Scientia Iranica}, 19(3):674--679, 2012.

\bibitem{mashhadi2012novel}
Samaneh Mashhadi.
\newblock A novel secure self proxy signature scheme.
\newblock {\em IJ Network Security}, 14(1):22--26, 2012.

\bibitem{sun2020toward}
Zhen Sun, Liyuan Song, Qin Huang, Liuguo Yin, Guilu Long, Jianhua Lu, and Lajos Hanzo.
\newblock Toward practical quantum secure direct communication: A quantum-memory-free protocol and code design.
\newblock {\em IEEE Transactions on Communications}, 68(9):5778--5792, 2020.

\bibitem{shi2017quantum}
Run-Hua Shi and Shun Zhang.
\newblock Quantum solution to a class of two-party private summation problems.
\newblock {\em Quantum Information Processing}, 16(9):1--9, 2017.

\bibitem{sutradhar2021efficient}
Kartick Sutradhar and Hari Om.
\newblock An efficient simulation for quantum secure multiparty computation.
\newblock {\em Scientific Reports}, 11(1):1--9, 2021.

\bibitem{sutradhar2020hybrid}
Kartick Sutradhar and Hari Om.
\newblock Hybrid quantum protocols for secure multiparty summation and multiplication.
\newblock {\em Scientific Reports}, 10(1):1--9, 2020.

\bibitem{shi2016quantum}
Run-hua Shi, Yi~Mu, Hong Zhong, Shun Zhang, and Jie Cui.
\newblock Quantum private set intersection cardinality and its application to anonymous authentication.
\newblock {\em Information Sciences}, 370:147--158, 2016.

\bibitem{shi2016comment}
Run-hua Shi, Yi~Mu, Hong Zhong, and Shun Zhang.
\newblock Comment on “secure quantum private information retrieval using phase-encoded queries”.
\newblock {\em Physical Review A}, 94(6):066301, 2016.

\bibitem{shi2018efficient}
Run-Hua Shi.
\newblock Efficient quantum protocol for private set intersection cardinality.
\newblock {\em IEEE Access}, 6:73102--73109, 2018.

\bibitem{Qin2018Multidimensional}
Huawang Qin, Raylin Tso, and Yuewei Dai.
\newblock Multi-dimensional quantum state sharing based on quantum fourier transform.
\newblock {\em Quantum Information Processing}, 17(3):48, 2018.

\bibitem{bao2009threshold}
Li~Bao-Kui, Yang Yu-Guang, and Wen Qiao-Yan.
\newblock Threshold quantum secret sharing of secure direct communication.
\newblock {\em Chinese Physics Letters}, 26(1):010302, 2009.

\bibitem{sutradhar2021cost}
Kartick Sutradhar and Hari Om.
\newblock A cost-effective quantum protocol for secure multi-party multiplication.
\newblock {\em Quantum Information Processing}, 20:1--10, 2021.

\bibitem{yang2013secret}
Wei Yang, Liusheng Huang, Runhua Shi, and Libao He.
\newblock Secret sharing based on quantum fourier transform.
\newblock {\em Quantum information processing}, 12(7):2465--2474, 2013.

\bibitem{mashhadi2017new}
Samaneh Mashhadi.
\newblock New multi-stage secret sharing in the standard model.
\newblock {\em Information Processing Letters}, 127:43--48, 2017.

\bibitem{hillery1999quantum}
Mark Hillery, Vladim{\'\i}r Bu{\v{z}}ek, and Andr{\'e} Berthiaume.
\newblock Quantum secret sharing.
\newblock {\em Physical Review A}, 59(3):1829, 1999.

\bibitem{eastlake2001us}
Donald Eastlake and Paul Jones.
\newblock Us secure hash algorithm 1 (sha1), 2001.

\bibitem{CommentKao2018}
Shih-Hung Kao and Tzonelih Hwang.
\newblock Comment on (t, n) threshold d-level quantum secret sharing.
\newblock {\em arXiv preprint arXiv:1803.00216}, 2018.

\bibitem{mashhadi2020improvement}
Samaneh Mashhadi.
\newblock Improvement of a (t, n) threshold d- level quantum secret sharing scheme.
\newblock {\em Journal of Applied Security Research}, pages 1--12, 2020.

\bibitem{shamir1979share}
Adi Shamir.
\newblock How to share a secret.
\newblock {\em Communications of the ACM}, 22(11):612--613, 1979.

\bibitem{sutradhar2023quantum}
Kartick Sutradhar, Ranjitha Venkatesh, and Priyanka Venkatesh.
\newblock Quantum internet of things for smart healthcare.
\newblock In {\em Learning Techniques for the Internet of Things}, pages 261--285. Springer, 2023.

\bibitem{sutradhar2021secret}
Kartick Sutradhar and Hari Om.
\newblock Secret sharing based multiparty quantum computation for multiplication.
\newblock {\em International Journal of Theoretical Physics}, 60(9):3417--3425, 2021.

\bibitem{sutradhar20213efficient}
Kartick Sutradhar and Hari Om.
\newblock Efficient cryptographic protocol for sorting with data-oblivious.
\newblock In {\em 2021 2nd International Conference for Emerging Technology (INCET)}, pages 1--6. IEEE, 2021.

\bibitem{sutradhar20244quantum}
Kartick Sutradhar, Ranjitha Venkatesh, and Priyanka Venkatesh.
\newblock Quantum blockchain-based healthcare: Merging frontiers for secure and efficient data management.
\newblock In {\em Healthcare Services in the Metaverse}, pages 190--207. CRC Press, 2024.

\bibitem{sutradhar2024smart}
Kartick Sutradhar, Ranjitha Venkatesh, and Priyanka Venkatesh.
\newblock Smart healthcare services employing quantum internet of things on metaverse.
\newblock In {\em Healthcare Services in the Metaverse}, pages 170--189. CRC Press, 2024.

\bibitem{challagundla2024privacy}
Koushik Challagundla and Kartick Sutradhar.
\newblock A privacy-preserving quantum authentication for vehicular communication.
\newblock {\em Quantum Information Processing}, 23(11):1--24, 2024.

\bibitem{sutradhar2024review}
Kartick Sutradhar, Ranjitha Venkatesh, and Priyanka Venkatesh.
\newblock A review on smart healthcare employing quantum internet of things.
\newblock {\em IEEE Engineering Management Review}, 2024.

\bibitem{challagundla2024secure}
Koushik Challagundla and Kartick Sutradhar.
\newblock A secure quantum protocol for vehicular ad hoc networks.
\newblock In {\em 2024 15th International Conference on Computing Communication and Networking Technologies (ICCCNT)}, pages 1--6. IEEE, 2024.

\bibitem{nayaka2024survey}
Parikshith Nayaka Sheetakallu~Krishnaiah, Dayanand~Lal Narayan, and Kartick Sutradhar.
\newblock A survey on secure metadata of agile software development process using blockchain technology.
\newblock {\em Security and Privacy}, 7(2):e342, 2024.

\bibitem{sutradhar2024survey}
Kartick Sutradhar, Beena~G Pillai, Ruhul Amin, and Dayanand~Lal Narayan.
\newblock A survey on privacy-preserving authentication protocols for secure vehicular communication.
\newblock {\em Computer Communications}, 2024.

\bibitem{sutradhar2023secure}
Kartick Sutradhar.
\newblock Secure multiparty quantum aggregating protocol.
\newblock {\em Quantum Inf. Comput.}, 23(3\&4):245--256, 2023.

\bibitem{sutradhar2024svqcp}
Kartick Sutradhar and Ranjitha Venkatesh.
\newblock Svqcp: A secure vehicular quantum communication protocol.
\newblock {\em IEEE Transactions on Network Science and Engineering}, 2024.

\bibitem{sutradhar2022privacy}
Kartick Sutradhar and Hari Om.
\newblock A privacy-preserving comparison protocol.
\newblock {\em IEEE Transactions on Computers}, 72(6):1815--1821, 2022.

\bibitem{sutradhar20211efficient}
Kartick Sutradhar and Hari Om.
\newblock An efficient simulation of quantum secret sharing.
\newblock {\em arXiv preprint arXiv:2103.11206}, 2021.

\bibitem{sutradhar2024privacy}
Kartick Sutradhar and Ranjitha Venkatesh.
\newblock A privacy preserving quantum aggregating technique.
\newblock {\em Quantum Information Processing}, 23(4):124, 2024.

\bibitem{sutradhar20244privacy}
Kartick Sutradhar and Ranjitha Venkatesh.
\newblock A privacy preserving quantum aggregating technique with simulation.
\newblock {\em Physica Scripta}, 99(5):055105, 2024.

\bibitem{shi2016secure}
Run-hua Shi, Yi~Mu, Hong Zhong, Jie Cui, and Shun Zhang.
\newblock Secure multiparty quantum computation for summation and multiplication.
\newblock {\em Scientific reports}, 6(1):1--9, 2016.

\end{thebibliography}
%\bibliographystyle{unsrt}

\end{document}